# Deciphering chemical order/disorder and material properties at the single-atom level

Yongsoo Yang[1*], Chien-Chun Chen[1,2*], M. C. Scott[1,3*], Colin Ophus[3*], Rui Xu[1], Alan Pryor Jr[1], Li Wu[1], Fan Sun[4], W. Theis[5], Jihan Zhou[1], Markus Eisenbach[6], Paul R. C. Kent[7,8], Renat F. Sabirianov[9], Hao Zeng[4], Peter Ercius[3] & Jianwei Miao[1†]

*1Department of Physics & Astronomy and California NanoSystems Institute, University of California, Los Angeles, CA 90095, USA. 2Department of Physics, National Sun Yat-sen University, Kaohsiung 80424, Taiwan. 3National Center for Electron Microscopy, Molecular Foundry, Lawrence Berkeley National Laboratory, Berkeley, CA 94720, USA. 4Department of Physics, University at Buffalo, the State University of New York, Buffalo, NY 14260, USA. 5Nanoscale Physics Research Laboratory, School of Physics and Astronomy, University of Birmingham, Edgbaston, Birmingham B15 2TT, UK. 6National Center for Computational Sciences, Oak Ridge National Laboratory, Oak Ridge, TN 37831, USA. 7Computer Science and Mathematics Division, Oak Ridge National Laboratory, Oak Ridge, TN 37831, USA. 8Center for Nanophase Materials Sciences, Oak Ridge National Laboratory, Oak Ridge, TN 37831, USA. 9Department of Physics, University of Nebraska at Omaha, Omaha, NE 68182, USA.*

*These authors contributed equally to this work. †Email: miao@physics.ucla.edu*

**Correlating 3D arrangements of atoms and defects with material properties and functionality forms the core of several scientific disciplines. Here, we determined the 3D coordinates of 6,569 iron and 16,627 platinum atoms in a model iron-platinum nanoparticle system to correlate 3D atomic arrangements and chemical order/disorder with material properties at the single-atom level. We identified rich structural variety and chemical order/disorder including 3D atomic composition,**

**grain boundaries, anti-phase boundaries, anti-site point defects and swap defects. We show for the first time that experimentally measured 3D atomic coordinates and chemical species with 22 pm precision can be used as direct input for first-principles calculations of material properties such as atomic magnetic moments and local magnetocrystalline anisotropy. This work not only opens the door to determining 3D atomic arrangements and chemical order/disorder of a wide range of nanostructured materials with high precision, but also will transform our understanding of structure-property relationships at the most fundamental level.**

Perfect crystals are rare in nature. Real materials are often composed of crystal defects and chemical order/disorder such as grain boundaries, dislocations, interfaces, surface reconstructions and point defects that disrupt the periodicity of the atomic arrangement and determine their properties and performance[1-5]. One prominent example is intermetallic compounds involving two or more atomic species, in which chemical order/disorder determines their mechanical, catalytic, optical, electronic and magnetic properties[6-9]. For instance, as-synthesized at room temperature, FePt nanoparticles and thin films with a near-1:1 composition have a chemically disordered face-centered cubic (fcc) structure (A1 phase)[7,10,11]. When annealed at high temperatures, they undergo a transformation from an A1 phase to an ordered face-centered tetragonal (L1$_0$) phase[7,10,11]. Due to the chemical ordering and strong spin-orbit coupling[12], L1$_0$ FePt exhibits extremely large magnetocrystalline anisotropy energy (MAE) and is among the most promising candidates for next-generation magnetic storage media[11-13] and permanent magnet applications[14]. Recently, it has also been reported that L1$_0$ ordered FePt nanoparticles show enhanced catalytic properties for oxygen evolution reaction[15]. However, although this material system has attracted considerable attention, a fundamental understanding of 3D chemical order/disorder, crystal defects and its magnetic properties at the individual atomic level remains elusive.

On a parallel front, quantum mechanics calculations such as density functional theory (DFT) have made significant progress from modelling ideal bulk systems to "real" materials with dopants, dislocations, grain boundaries and interfaces[16,17]. Presently, these calculations rely heavily on average atomic models extracted from crystallography. However, for complex materials such as non-stoichiometric compounds, materials deviating from thermodynamic equilibrium, or materials subject to complex processing treatments, it is either impossible or hugely computationally expensive to perform structural optimization using *ab initio* methods[16-19]. To improve the predictive power of DFT calculations, there is a pressing need to provide atomic coordinates of real systems beyond average crystallographic measurements as input for computation. One powerful method to address this challenge is atomic electron tomography (AET)[20-24], in which the 3D positions of individual atoms with defects can be measured without assuming crystallinity or using averaging[25-28].

Here, we report the precise determination of the 3D coordinates and chemical species of 23,196 atoms in a single 8-nm $Fe_{0.28}Pt_{0.72}$ nanoparticle using a generalized Fourier iterative reconstruction (GENFIRE) algorithm. From these atomic coordinates and species, we revealed chemical order/disorder and crystal defects such as grain boundaries, anti-phase boundaries, anti-site and swap defects in 3D. Furthermore, the measured atomic coordinates were used as direct input for DFT calculations to correlate 3D chemical ordering and defects with the local MAE. This work marries the forefront of 3D atomic structure determination of crystal defects and chemical order/disorder with DFT calculations, which we expect to find broad applications in physics, chemistry, materials science, nanoscience and nanotechnology.

**3D determination of atomic positions and species**

FePt nanoparticles with a size of ~8nm were synthesized by an organic solution phase synthesis technique[29] and annealed at 600°C for 25 minutes to induce partial chemical ordering (Methods). Using an aberration-corrected scanning transmission electron microscope (STEM) operated in annular dark-field (ADF) mode[30-32], we acquired high-resolution tomographic tilt series from several FePt nanoparticles. A representative tilt series of 68 images with a tilt range from -65.6° to +64.0° was chosen for the detailed analysis due to its structural complexity (Extended Data Fig. 1). After being denoised[33] (Methods, Extended Data Fig. 2) and aligned[21], the tilt series was used to compute a 3D reconstruction with a newly developed GENFIRE algorithm (Methods). GENFIRE started with a 3D regular grid in Fourier space and assigned some grid points with magnitudes and phases derived from the experimental tilt series through oversampling[34,35] and the discrete Fourier transform. The assembled 3D grid consists of a fraction of grid points assigned with measured data, while the remaining grid points were set as undefined. The algorithm then iterated between real and reciprocal space using the fast Fourier transform and its inverse. A support and positivity were used as real space constraints, and the measured grid points were enforced in reciprocal space, while the undefined points were iteratively updated by the algorithm. After 500 iterations, a converged 3D reconstruction was obtained. Next, we refined each tilt angle by projecting the 3D reconstruction to calculate a 2D image (Methods). An error metric was computed between the calculated and measured images. A more accurate tilt angle was obtained by optimizing the orientation of the 3D reconstruction to produce a calculated projection image and minimize the error metric. After refining all the tilt angles, we repeated the GENFIRE reconstruction and produced a final 3D reconstruction. Compared to equal slope tomography[36] that has been applied to AET[21,23,25], GENFIRE not only produces more accurate 3D reconstructions through the refinement of the tilt angles, but also represents a more general method applicable to any tilt geometry and capable of solving larger systems in a faster manner.

From the final 3D reconstruction, we identified all local intensity maxima taking into account a minimum distance of 2Å between two neighbouring atoms, which is justified as the covalent diameter of an Fe atom is 2.52Å. Extended Data Fig. 3a shows the histogram of the local intensity maxima, indicating each local maximum belongs to one of the three categories: i) potential Pt atoms, ii) potential Fe atoms and iii) potential non-atoms (intensity too weak to be an atom). By developing an unbiased classification method (Methods, Extended Data Fig. 3), we classified each atom into one of these three categories. However, experimental data could not be measured beyond -65.6° and +64.0° (known as the missing wedge problem), resulting in lower average atom intensity in the missing wedge region. To mitigate this effect, we re-classified the local intensity maxima in this region. This atom tracing and classification process yielded 17,087 Pt and 6,717 Fe atom candidates. To validate the robustness of our method with regards to the choice of the minimum distance, we changed the minimum distance to 1.6Å and repeated the whole process, producing 16,551 Pt and 6,639 Fe atom candidates. The majority of the atom candidates between the two independent sets are consistent and identified as real atoms. A very small fraction of inconsistent atom candidates were validated one-by-one using the 68 experimental images to produce a 3D atomic model (Methods). The model was further refined using the experimental images to improve the precision of individual atom coordinates[25] (Extended Data Fig. 4), resulting in a final 3D model of 23,196 atoms, consisting of 16,627 Pt and 6,569 Fe atoms.

To verify the final atomic model, we applied multislice STEM simulations[37] to calculate 68 images from the model using the same experimental parameters. Extended Data Figs. 5a-c show good agreement between a measured and multislice image. Using the same reconstruction, atom tracing and classification procedures, we obtained a new 3D model consisting of 16,579 Pt and 6,791 Fe atoms. Compared to the experimental final model, 99.0% of all atoms are correctly identified in the new model and a root-mean-square deviation of the common atom positions is 22 pm (Extended Data Fig. 5d). To further confirm the precision of our atomic position measurements, we performed a lattice analysis of the 3D atomic model (Extended Data Figs. 6) and determined the 3D atomic displacements of the nanoparticle (Extended Data Figs. 7). Based on the atomic displacements relative to an ideal fcc lattice, we estimated an average 3D precision of 21.6 pm (Extended Data Fig. 8a), which agrees well with the multislice result.

**3D identification of chemical order/disorder**

From the 3D positions of individual atomic species (Fig. 1a), we classified the 3D chemical order/disorder of the FePt nanoparticle. This was achieved by determining the short-range order parameter (SROP) of all phases present in the 3D structure (Methods). The nanoparticle consists of two large chemically ordered fcc ($L1_2$) $FePt_3$ grains that form concave shapes and are nearly connected (Fig. 1b). Seven smaller grains are located at the boundary between the two large $L1_2$ grains or near the centre of the nanoparticle, including three $L1_2$ $FePt_3$ grains, three $L1_0$ FePt grains and a Pt-rich A1 grain (Fig. 1b and Supplementary Video 1). This level of complexity in the 3D structure and chemical order/disorder can only be fully revealed by AET. To illustrate this point, we used multislice STEM simulations to calculate 2D projection images from the 3D atomic model along the [100], [010] and [001] directions (Fig. 1c). Although 2D analysis of the images indicates that the nanoparticle is primarily composed of $L1_2$ $FePt_3$ grains, several of the 'L1$_0$ grain' signatures appearing in the projection images (magenta in Fig. 1c) are deceptive structural information, as these derive from the overlapping projections of the two $L1_2$ grains, rather than from actual $L1_0$ grains.

Next, we analysed the chemical order/disorder associated with the 3D grains. Figure 2a shows the atomic positions and chemical species of the nanoparticle with grain boundaries marked as black lines. The grains identified are more ordered in their cores and less ordered closer to the grain surface. Figures 2b-e show four representative cut-outs of the atomic model where the SROP of the corresponding phase is averaged along the [010] direction and displayed as the background colour. The most highly chemically ordered region of the nanoparticle is at the core of a large $L1_2$ grain with the SROP close to 1 (Fig. 2b). Figure 2c shows a grain boundary between two large $L1_2$ grains with a varied grain boundary width. Anti-phase boundaries between the two $L1_2$ grains are also observed (Extended Data Fig. 8b). The largest $L1_0$ grain is shown in Fig. 1b (the 3[rd] grain from the left) and Fig. 2d. The $L1_0$ ordering can be identified by the high concentration of Fe atoms on alternating planes along the vertical direction. This $L1_0$ grain sits between the two large $L1_2$ $FePt_3$ grains with each of its two Fe sublattices matching the Fe sublattice of one of the neighbouring $L1_2$ grains (Extended Data Fig. 6), suggesting the shared Fe planes with its neighbouring grains may have facilitated the nucleation of the $L1_0$ phase. The central region of the nanoparticle has the highest degree of chemical disorder, including a Pt-rich A1-phase grain (Fig. 2e), with much lower SROP values than those in the two large $L1_2$ grains.

To probe the 3D chemical order/disorder at the single-atom level, we analysed individual anti-site point defects in the 3D reconstruction of the nanoparticle. Figures 3a, b and Extended Data Fig. 8b show 3D atomic positions overlaid on the reconstructed intensity of several representative anti-site point defects (arrows) in the $L1_2$ $FePt_3$ grains, where an Fe atom occupies a Pt atom site or vice versa. A perfect $L1_2$ $FePt_3$ phase is illustrated in Fig. 3d for reference. As our 3D reconstruction does not use an *a priori* assumption of crystallinity, the identification of individual atomic species is based on the intensity distribution around local maxima of the 3D reconstruction (Extended Data Fig. 4). The anti-site point defects in these figures are clearly visible by comparing their local peak intensity with that of the surrounding Pt and Fe atoms. Furthermore, swap defects are also observed (Fig. 3c), where a pair of nearest-neighbour Fe and Pt atoms are swapped. Overall, the FePt nanoparticle contains a substantial number of anti-site defects and chemical disorder. Figures 3e and g show the anti-site defect density of the two large $L1_2$ grains (inset) as a function of the distance from the grain surface. Far outside of each grain, the anti-site defect density approaches ~50%. This is because two of the four sublattices in the two large $L1_2$ grains share the same composition (pure Pt), while the other two sublattices swap Fe for Pt and vice versa (Extended Data Fig. 6). The anti-site defect density drops to below 40% at the surface of the two grains and reduces to ~3% for sites deep inside each grain. Figures 3f and h show the SROP of the two large $L1_2$ grains as a function of the distance from the grain surface. The negative SROP values outside each grain indicate the local structure is anti-correlated with that of the other $L1_2$ grain.

The striking similarities between the two large L1$_2$ grains, *i.e.* each having a concave shape with a highly-ordered core, a similar chemically disordered boundary and a consistent distribution of the anti-site defect density (Figs. 3e-h), suggest a potential pathway in the nucleation and growth process. We note that as-synthesized FePt nanoparticles show large chemical disorder with a Pt-rich core[38]. Such a 3D Pt-rich core is observed in our measurement (Fig. 2e). During the annealing process, Pt atoms diffused out from the core[38] and the nucleation of the L1$_2$ phase likely occurred simultaneously at multiple sites in the nanoparticle. The nuclei then grew and merged into larger grains by the Ostwald ripening process[39]. This process would continue until the nanoparticle became a single crystal if sufficiently high temperature/long time annealing was applied. However, if the annealing process was stopped at some intermediate stage, two or more larger grains with similar sizes could coexist since it was difficult for either to annihilate the others. The chemical ordering at the grain boundaries would then be frustrated by competition between neighbouring grains. However, determining the particle's chemical structure growth pathway with certainty will require adding the dimension of time to the AET measurements. In principle, it is possible to measure a series of tomographic data sets of the same particle at different times during heat treatment. This would unlock the atomic-scale mechanism underlying the nucleation and growth process.

**Correlating chemical order/disorder to magnetic properties**

To relate experimentally determined atomic coordinates and chemical order/disorder to magnetic properties, we performed DFT calculations of the atomic magnetic moments and MAEs. We focused on one of the grain boundaries between two large L1$_2$ grains, where the largest L1$_0$ grain is located, and calculated the MAE using two independent approaches (Methods). First, we cut out a 1,470-atom supercell from this region, containing the largest $L1_0$ grain. By sliding a 32-atom-cubic volume in the supercell with a half-unit-cell per step along each direction, we obtained 1,452 32-atom volumes and calculated their local MAEs with DFT. Second, we cropped six nested supercells of the same region with a size ranging from 32 to 1,372 atoms and calculated the atomic magnetic moments (Extended Data Fig. 9) and MAEs. Figure 4a and Extended Data Fig. 10a show a good agreement of the MAEs between the supercell and sliding volume calculations, which validates the sliding volume approach for determining the local MAEs. The correlation between the MAE and supercell size was also reproduced by modelling a spherical-shaped $L1_0$ grain, enclosed by cubic $L1_2$ grains with different sizes. In this model, the MAE of the $L1_0$ grain was obtained from the 32-atom supercell calculation, and the radius of the $L1_0$ sphere and the MAE of the $L1_2$ grains were fitted to the MAEs obtained from six nested supercells (Figure 4a and Extended Data Fig. 10a). To study the influence of measured atomic positions, we self-consistently relaxed four 32-atom, one 256-atom and one 500-atom volumes using DFT. The root-mean-square deviation between the measured and relaxed atomic positions is 24.7 pm, which agrees with our precision estimation. The average MAE difference is 0.064 meV/atom, indicating that our measured atomic coordinates are sufficiently accurate for DFT calculations and the atomic composition and chemical order/disorder are the main determinants of the local MAE.

Figure 4b and Extended Data Fig. 10b show a strong correlation between the local MAEs of all sliding 32-atom volumes and the $L1_0$ order parameter difference, where the $L1_0$ order parameters were computed from the same 32-atom volumes. The 3D distribution of the local MAEs matches well with that of the $L1_0$ order parameter difference inside the 1,470-atom supercell (Fig. 4c and Extended Data Fig. 10c). The MAEs obtained by our sliding volume approach can be used to provide the local uniaxial anisotropy constants for micromagnetic simulations[40,41], which are presently taken from either bulk or modelled values without considering atomic details. Because there is no perfect $L1_0$ phase in the nanoparticle, the largest local MAE in the region (0.945 meV/atom) is smaller than that of an ideal $L1_0$ phase (1.40 meV/atom) obtained from our DFT calculation. The smallest MAEs exist in the $L1_2$ grain and some sharp transitions from the large and small MAEs are also observed. Fig. 4d and Extended Data Fig. 10d show the local MAE distribution at an $L1_0$ and $L1_2$ grain boundary, overlaid with measured atomic positions and species. The sharp grain boundary is responsible for a sudden transition of the local MAE. Although here we used an FePt model system to correlate 3D grain boundaries and chemical order/disorder with magnetic properties at the single-atom level, this method can be readily applied to many other material systems.

## Conclusions

We have developed a new tomographic reconstruction algorithm, termed GENFIRE, to determine the 3D coordinates and chemical species of 23,196 atoms in a FePt nanoparticle with 22 pm precision. We have identified atomic composition, chemical order/disorder, grain boundaries, anti-phase boundaries, anti-site point defects and swap defects with unprecedented 3D detail. Based on a statistical analysis of the chemical order/disorder and anti-site defect densities of the two large $L1_2$ grains, we suggested a potential pathway in the nucleation and growth process of the chemically ordered grains inside the nanoparticle. We also, for the first time, used the experimentally measured 3D atomic coordinates and species with defects as direct input for DFT calculations to correlate 3D chemical order/disorder and grain boundaries with magnetic properties at the single-atom level. This work makes significant advances in characterization capabilities and expands our fundamental understanding of structure-property relationships. As a general and powerful 3D reconstruction algorithm, GENFIRE can be broadly applied to other systems and tomographic fields. This work also lays the foundation for precisely determining the 3D atomic arrangement of chemical order/disorder of a wide range of nanostructured materials, especially those where conventional 2D projection images may provide deceptive structural information. Additionally, the ability to use measured atomic coordinates as input for DFT calculations to correlate 3D structure and chemical order/disorder with material properties is expected to have a profound impact across several disciplines. Finally, with further development, our method can in principle be applied to understand nucleation and growth mechanisms at the individual atomic level.

**Acknowledgements** We thank J. Shan, J. A. Rodriguez, M. Gallagher-Jones and J. Ma for their help with this project. This work was primarily supported by the Office of Basic Energy Sciences of the US DOE (DE-SC0010378). This work was partially supported by NSF (DMR-1437263) and ONR MURI (N00014-14-1-0675). The chemical ordering analysis and ADF-STEM imaging with TEAM I were performed at the Molecular Foundry, which is supported by the Office of Science, Office of Basic Energy

Sciences of the U.S. DOE under Contract No. DE-AC02—05CH11231. M.E. (DFT calculations) was supported by the U.S. DOE, Office of Science, Basic Energy Sciences, Material Sciences and Engineering Division. DFT calculations by P.K. were conducted at the Center for Nanophase Materials Sciences, which is a DOE Office of Science User Facility. This research used resources of the Oak Ridge Leadership Computing Facility, which is supported by the Office of Science of the U.S. DOE under contract DE-AC05-00OR22725.

**Figure legends**

**Figure 1**. **3D determination of atomic coordinates, chemical species and grain structure of an FePt nanoparticle. a**, Overview of the 3D positions of individual atomic species with Fe atoms in red and Pt atoms in blue. **b**, The nanoparticle consists of two large $L1_2$ grains, three small $L1_2$ grains, three small $L1_0$ grains and a Pt-rich A1 grain. **c**, Multislice projection images obtained from the experimental 3D atomic model along the [100], [010] and [001] directions, where several '$L1_0$ grains' (magenta) appearing in the 2D images are deceptive structural information, as these derive from the overlapping projections of the two $L1_2$ grains. Scale bar, 2 nm.

**Figure 2**. **3D identification of grain boundaries and chemical order/disorder. a**, Atomic coordinates and species of the FePt nanoparticle divided into one-fcc-unit-cell thick slices. The grain boundaries are marked with black lines. **b-e**, Four representative cut-outs of the experimental atomic model, showing the most highly chemically ordered $L1_2$ region of the particle (**b**), a grain boundary between the two large $L1_2$ grains (**c**), the largest $L1_0$ grain (**d**), and the most chemically disordered region of the particle centered on a Pt-rich A1 grain (**e**). The locations of the cut-outs are labelled in (**a**), and the SROP of each cut-out is averaged along the [010] viewing direction and displayed as the background colour.

**Figure 3. Observation of anti-site point defects and swap defects and statistical analysis of the chemical order/disorder and anti-site density.** 3D atomic positions overlaid on the 3D reconstructed intensity illustrating anti-site point defects: a Pt atom occupying an Fe atom site (**a**), an Fe atom occupying a Pt atom site (**b**), a pair of nearest-neighbouring Fe and Pt atoms are swapped (swap defect) (**c**). **d**, 3D atomic structure of an ideal L1$_2$ FePt$_3$ phase for reference. **e, f**, The anti-site defect density and SROP for a large L1$_2$ grain, inset in (**e**), as a function of the distance from the grain surface (unit cell size = 3.875Å). **g, h**, The anti-site defect density and SROP for the other large L1$_2$ grain, inset in (**g**), as a function of the distance from the grain surface. Smooth red trendlines are overlaid on the defect density distribution as a guide for the eye.

**Figure 4. Local MAEs between the [100] and [001] directions determined by using measured atomic coordinates and species as direct input to DFT.** **a**, Black dots represent the MAEs calculated from six nested cubic volumes of 32, 108, 256, 500, 864 and 1,372 atoms. Blue curve shows the results of fitting a L1$_0$ sphere inside cubic L1$_2$ grains with different sizes. Red dots are the local MAEs averaged by sliding a 32-atom volume inside the corresponding six supercells. **b**, MAEs of all sliding 32-atom volumes as a function of the L1$_0$ order parameter difference of the same volume between the [100] and [001] directions. Negative MAE values indicate that their local magnetic easy axis is along the [100] instead of [001] direction. **c**, 3D iso-surface rendering of the local MAE (top) and L1$_0$ order parameter differences (bottom) inside the 1,470-atom supercell. **d**, Local MAE distribution at an L1$_0$ and L1$_2$ grain boundary, interpolated from the sliding local volume calculations and overlaid with measured atomic positions.

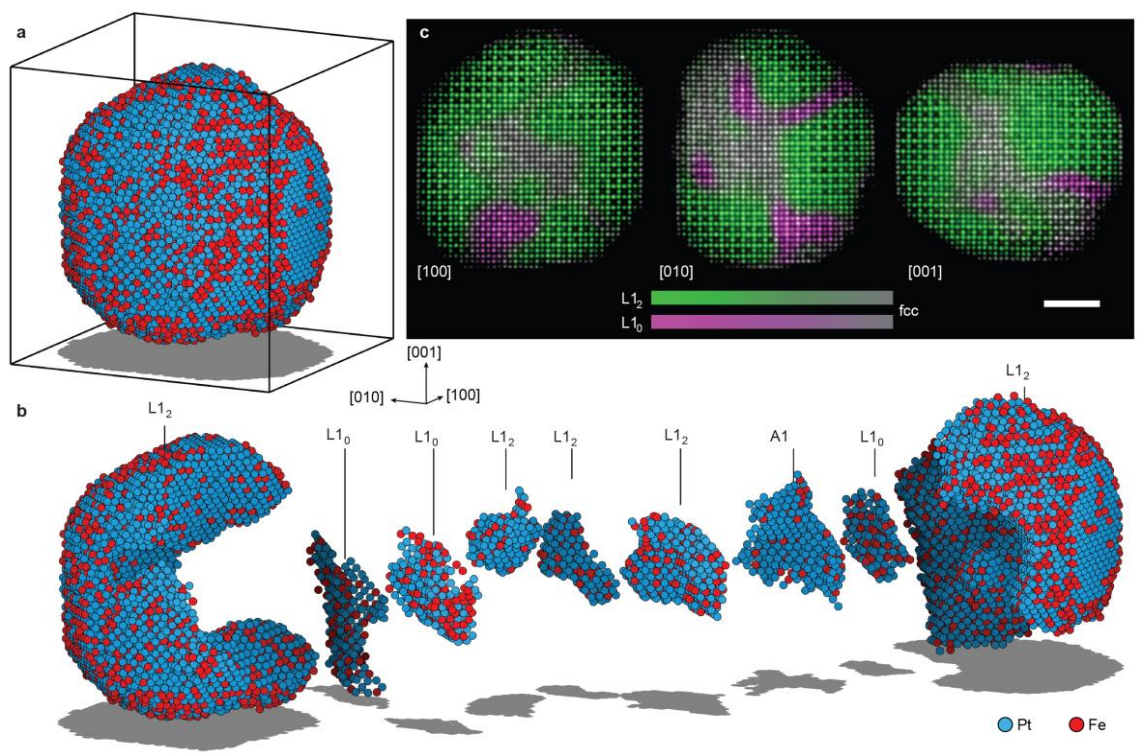

**Figure 1**

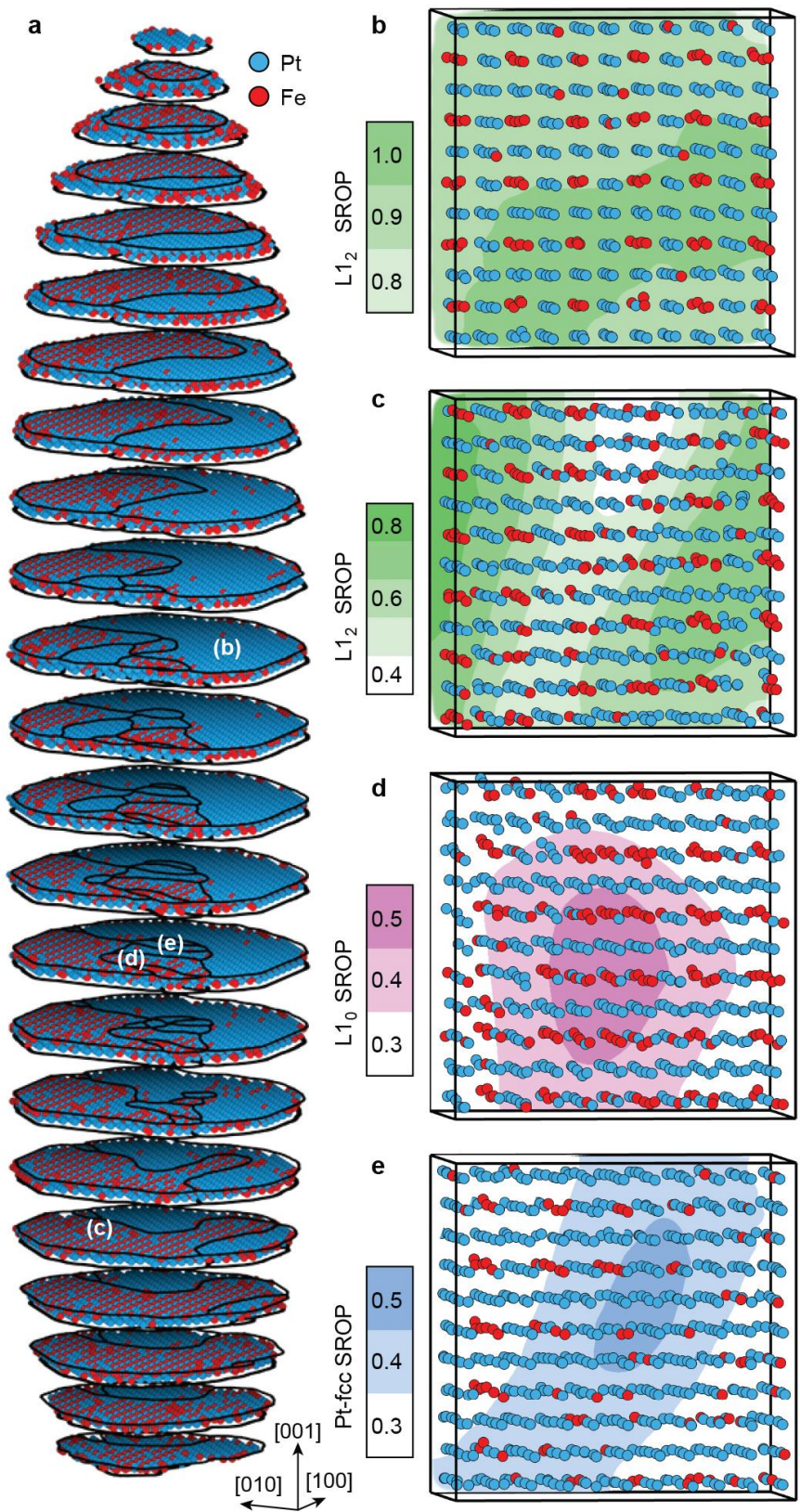

**Figure 2**

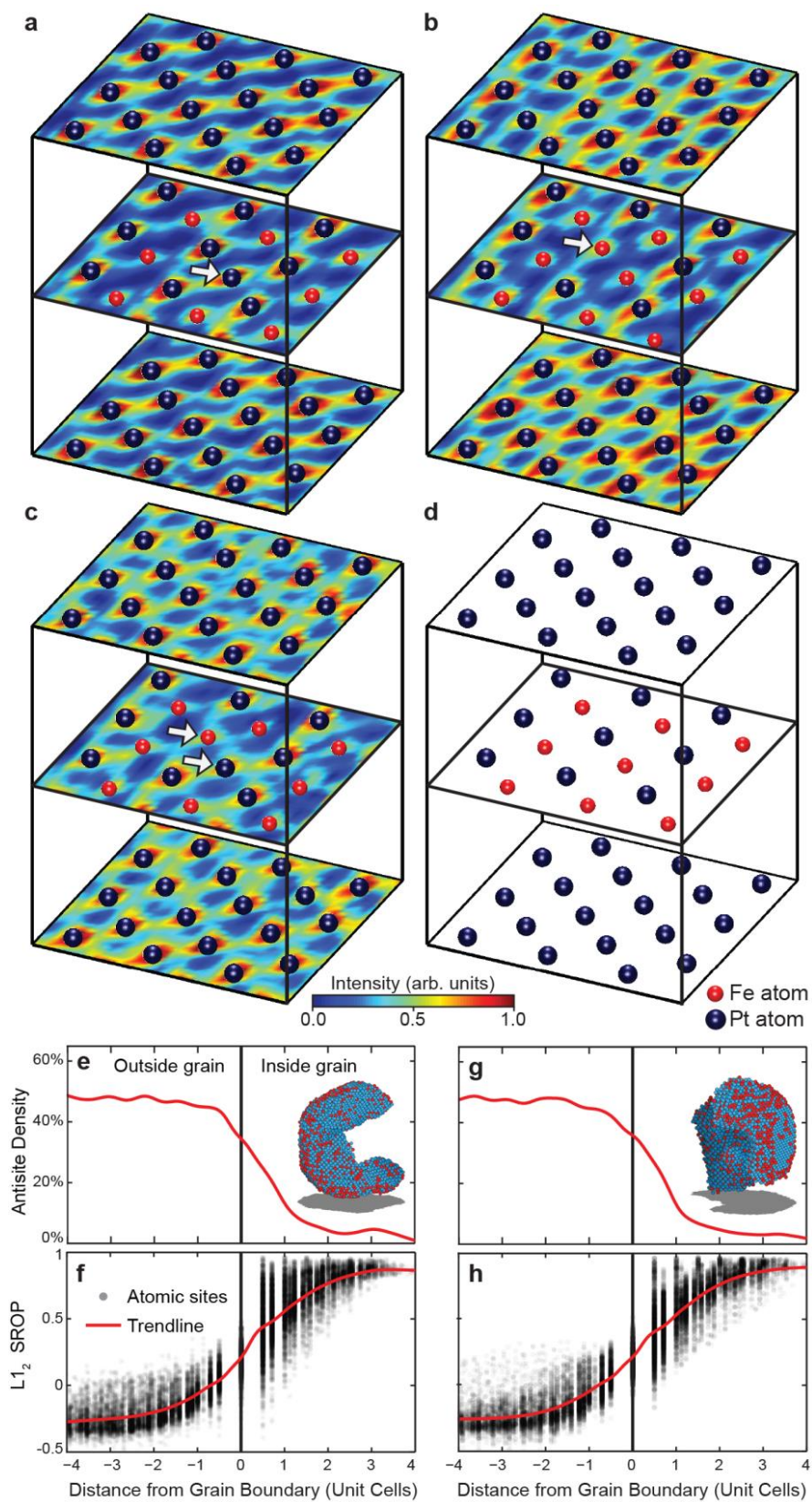

**Figure 3**

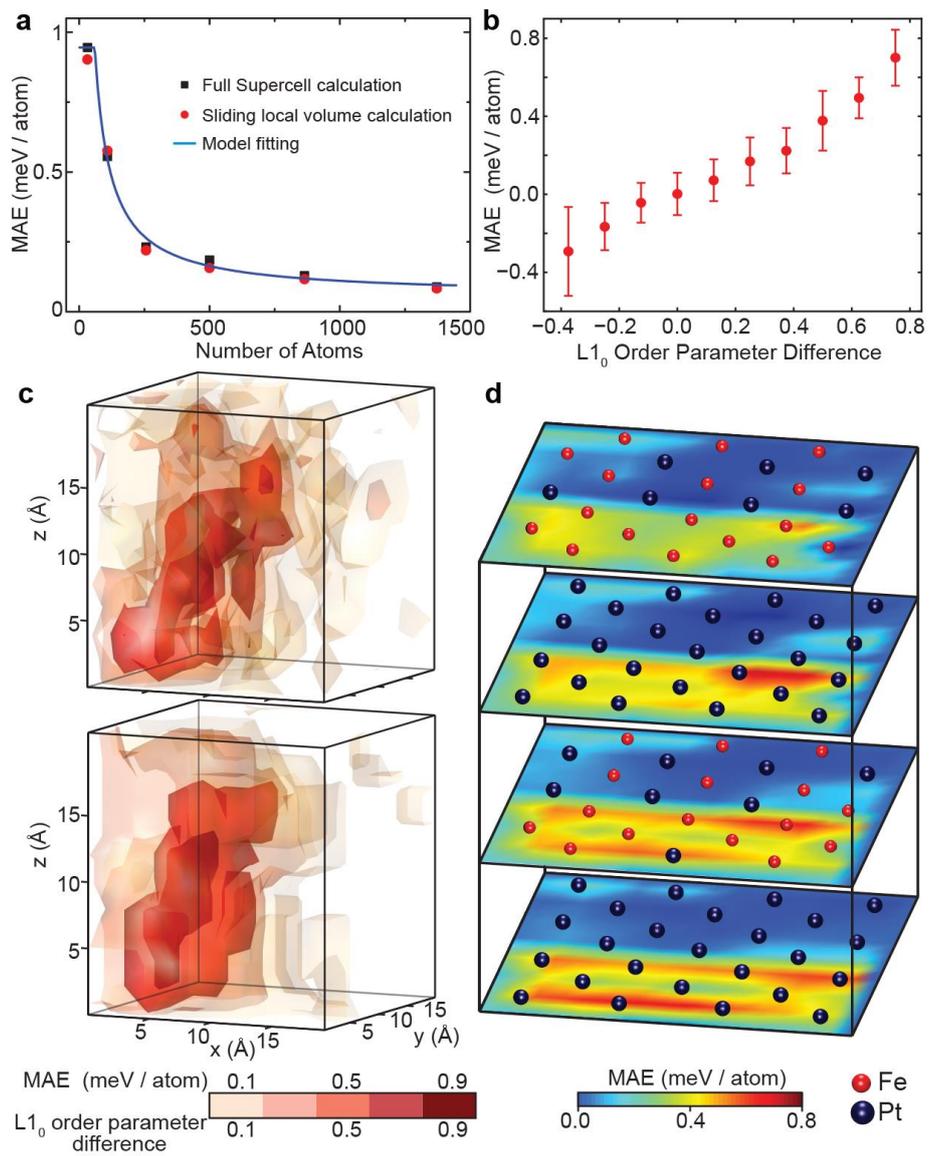

**Figure 4**